\documentclass{article}

\usepackage{PRIMEarxiv}

\usepackage[utf8]{inputenc} 
\usepackage[T1]{fontenc}    
\usepackage{cite}
\usepackage{amsmath,amssymb,amsfonts}
\usepackage{algorithmic}
\usepackage{graphicx}
\usepackage{textcomp}
\usepackage{siunitx}
\usepackage{caption}
\usepackage{subcaption}
\usepackage{booktabs}
\usepackage{multirow}
\usepackage{multicol}
\usepackage{wrapfig}
\usepackage{balance}
\usepackage{xcolor}
\usepackage{soul}
\usepackage{xspace}
\usepackage{physics}
\usepackage{url}
\usepackage{pifont}
\usepackage{threeparttable}
\usepackage{hyperref}

\usepackage[square, numbers, sort&compress]{natbib}
\newcommand{\etal}{\textit{et~al.}\xspace}
\newcommand{\cmark}{\ding{51}}%
\newcommand{\xmark}{\ding{55}}%
\graphicspath{{media/}}     

\title{Intra-video Positive Pairs in Self-Supervised Learning for Ultrasound
}

\author{
  Blake VanBerlo \\
  Cheriton School of Computer Science \\
  University of Waterloo \\
  Waterloo, Canada \\
  \texttt{bvanberl@uwaterloo.ca} \\
   \And
  Alexander Wong \\
  Department of Systems Design Engineering \\
  University of Waterloo \\
  Waterloo, Canada \\
   \And
  Jesse Hoey \\
  Cheriton School of Computer Science \\
  University of Waterloo \\
  Waterloo, Canada \\
     \And
  Robert Arntfield \\
  Division of Critical Care Medicine \\
  Western University \\
  London, Canada \\
}

\begin{document}
\maketitle

\begin{abstract}
Self-supervised learning (SSL) is one strategy for addressing the paucity of labelled data in medical imaging by learning representations from unlabelled images.
Contrastive and non-contrastive SSL methods produce learned representations that are similar for pairs of related images.
Such pairs are commonly constructed by randomly distorting the same image twice.
The videographic nature of ultrasound offers flexibility for defining the similarity relationship between pairs of images.
In this study, we investigated the effect of utilizing proximal, distinct images from the same B-mode ultrasound video as pairs for SSL.
Additionally, we introduced a sample weighting scheme that increases the weight of closer image pairs and demonstrated how it can be integrated into SSL objectives.
Named \textit{Intra-Video Positive Pairs} (IVPP), the method surpassed previous ultrasound-specific contrastive learning methods' average test accuracy on COVID-19 classification with the POCUS dataset by $\ge 1.3\%$.
Detailed investigations of IVPP's hyperparameters revealed that some combinations of IVPP hyperparameters can lead to improved or worsened performance, depending on the downstream task.
Guidelines for practitioners were synthesized based on the results, such as the merit of IVPP with task-specific hyperparameters, and the improved performance of contrastive methods for ultrasound compared to non-contrastive counterparts.
\end{abstract}


\section{Introduction}
\label{sec:introduction}

Medical ultrasound is a modality of imaging that uses the amplitude of ultrasonic reflections from tissues to compose a pixel map.
With the advent of point-of-care ultrasound devices, ultrasound has been increasingly applied in a variety of diagnostic clinical settings, such as emergency care, intensive care, oncology, and sports medicine~\cite{whitson2016ultrasonography,lau2022point,soni2019point,sood2019ultrasound,yim2012ultrasound}.
It possesses several qualities that distinguish it from other radiological modalities, including its portability, lack of ionizing radiation, and affordability.
Despite morphological distortion of the anatomy, ultrasound has been shown to be comparable to radiological alternatives, such as chest X-ray and CT, for several diagnostic tasks~\cite{van2011comparison,Alrajhi2012,Nazerian2015}.

Deep learning has been extensively studied as a means to automate diagnostic tasks in ultrasound.
As with most medical imaging tasks, the lack of open access to large datasets is a barrier to the development of such systems, since large training sets are required for deep computer vision models.
Organizations that have privileged access to large datasets are also faced with the problem of labelling ultrasound data.
Indeed, many point-of-care ultrasound examinations in acute care settings are not archived or documented~\cite{hall2016use,kessler2016effect}.

When unlabelled examinations are abundant, researchers turn to unsupervised representation learning to produce pretrained deep learning models that can be fine-tuned using labelled data.
Self-supervised learning (SSL) is a broad category these methods that has been explored for problems in diagnostic ultrasound imaging.
Broadly, SSL refers to the supervised pre-training of a machine learning model for a task that does not require labels for the task of interest.
The pre-training task (i.e., \textit{pretext task}) is a supervised learning task where the target is a quantity that is computed from unlabelled data.
After optimizing the model's performance on the pretext task, the weights are recast as initial weights for a new model that is trained to solve the task of interest (referred to as the \textit{downstream task}).
If the pretrained model has learned to produce representations of salient information in ultrasound images, then it is likely that it can be fine-tuned to perform the downstream task more proficiently than had it been randomly initialized.
Contrastive learning is a type of pretext task in SSL that involves predicting whether two inputs are related (i.e., positive pairs) or unrelated (i.e., negative pairs).
In computer vision, a common way to define positive pairs is to apply two randomly defined transformations to an image, producing two distorted views of the image with similar content.
Positive pairs satisfy a \textit{pairwise relationship} that indicate semantic similarity.
All other pairs of images are regarded as negative pairs. 
Non-contrastive methods disregard negative pairs, focusing only on reducing the differences between representations of positive pairs.

\begin{figure}
  \centering
  \begin{subfigure}{0.5\linewidth}
    \includegraphics[width=\linewidth]{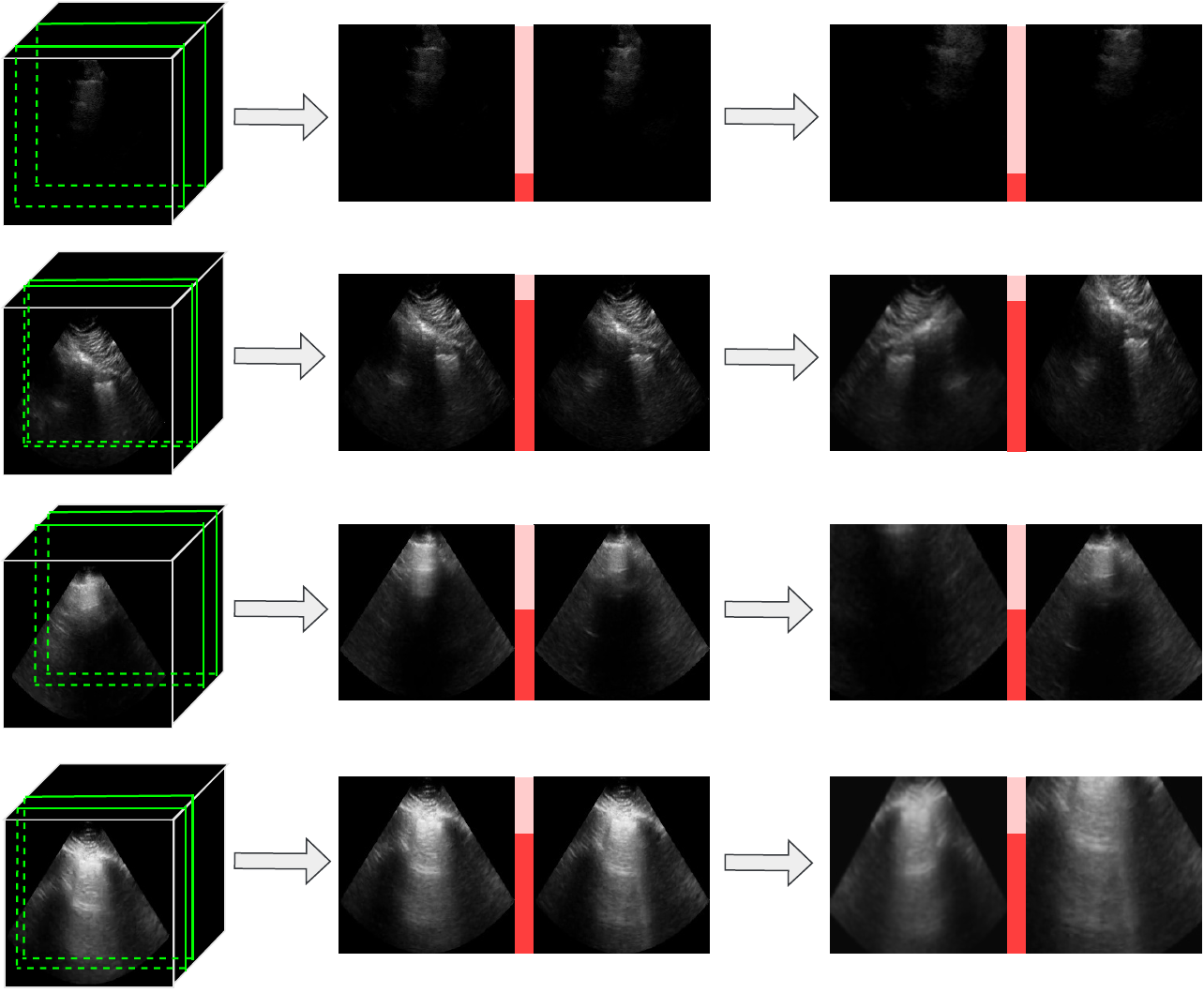}
    \caption{For B-mode ultrasound, positive pairs are B-mode images from the same video that are temporally separated.}
    \label{subfig:overview_bmode}
  \end{subfigure}
  \hfill
  \begin{subfigure}{0.49\linewidth}
    \centering
    \includegraphics[width=0.7\linewidth]{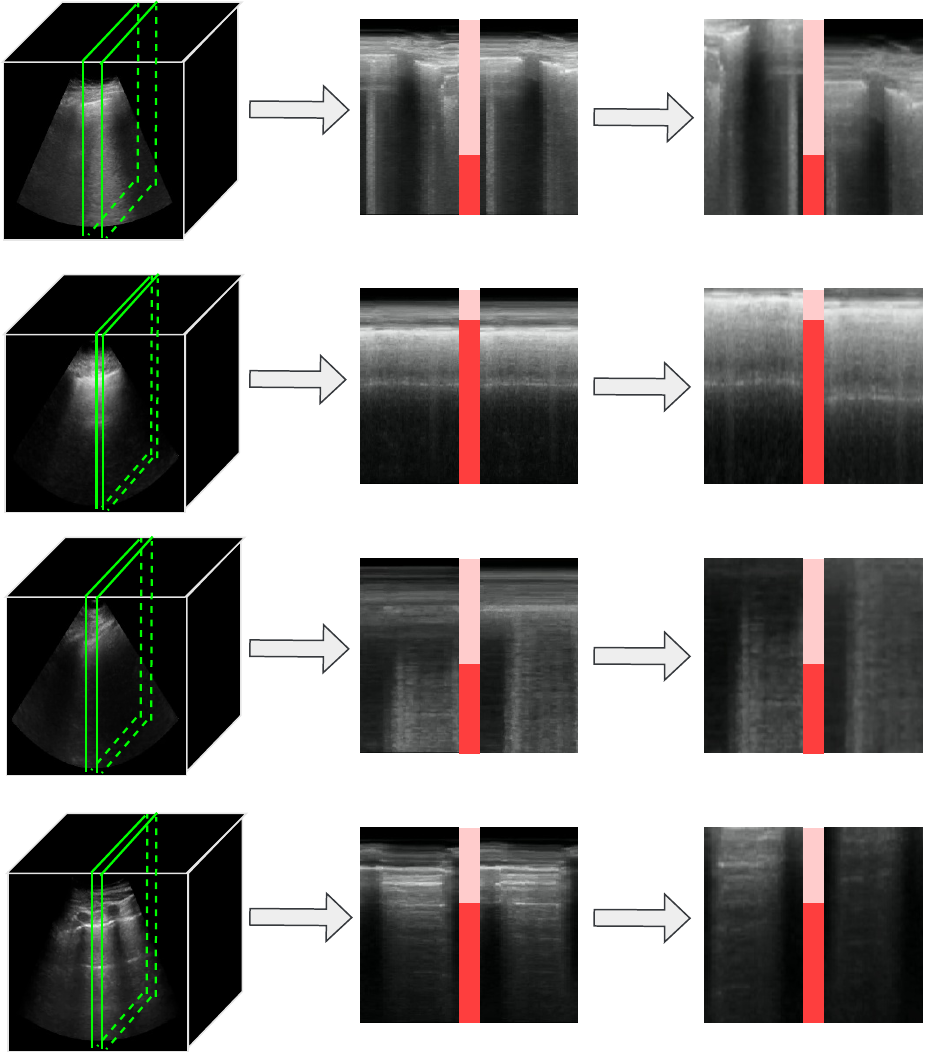}
    \vspace{5px}
    \caption{For M-mode ultrasound, positive pairs are M-mode images from the same B-mode video that are spatially separated.}
    \label{subfig:overview_mmode}
  \end{subfigure}
  \caption{
  An overview of the methods introduced in this study.
  Positive pairs of images for self-supervised learning (SSL) with ultrasound are randomly sampled from the same B-mode video, such that the distance between them does not exceed a threshold.
  Each image is then subjected to stochastic data augmentation.
  As indicated by the darkly shaded fraction of the red bar between pairs, sample weights that are inversely proportional to the distance between eah pair may be integrated into the SSL objective.
  }
  \label{fig:overview}
\end{figure}

Unlike other forms of medical imaging, US is a dynamic modality acquired as a stream of frames, resulting in a video.
Despite this, there are several US interpretation tasks that can be performed by assessing a still US image.
Previous studies exploring SSL in US have exploited the temporal nature of US by defining contrastive learning tasks with \textit{intra-video positive pairs} -- positive pairs comprised of images derived from the same video~\cite{chen2021uscl,basu_unsupervised_2022}.
Recent theoretical results indicate that the pairwise relationship must align with the labels of the downstream task in order to guarantee that self-supervised pretraining leads to non-inferior performance on the downstream task~\cite{Balestriero2022}.
For classification tasks, this means that positive pairs must have the same class label.
Due to the dynamic nature of US, one cannot assume that all frames in a US video possess the same label for all downstream US interpretation tasks.
As a result, it may be problematic to indiscriminately designate any pair of images originating from the same US video as a positive pair.
Moreover, since US videos are often taken sequentially as a part of the same examination or from follow-up studies of the same patient, different US videos may bear a striking resemblance to each other.
It follows that designating images from different US videos as negative pairs may result in negative pairs that closely resemble positive pairs.

In this study, we aimed to examine the effect of proximity and sample weighting of intra-video positive pairs for common SSL methods.
We also intended to determine if non-contrastive methods are more suitable for classification tasks in ultrasound.
Since non-contrastive methods do not require the specification of negative pairs, we conjectured that non-contrastive methods would alleviate the issue of cross-video similarity and yield stronger representations for downstream tasks.
%
Our contributions and results are summarized as follows:
\begin{itemize}
    \item A method for sampling intra-video positive pairs for joint embedding SSL with ultrasound.  
    \item A sample weighting scheme for joint embedding SSL methods that weighs positive pairs according to the temporal or spatial distance between them in their video of origin
    \item A comprehensive assessment of intra-video positive pairs integrated with SSL pretraining methods, as measured by downstream performance in B-mode and M-mode lung US classification tasks.
    We found that, with proper downstream task-specific hyperparameters, intra-video positive pairs can improve performance compared to the standard practice of producing two distortions of the same image.
    \item An comparison of contrastive versus non-contrastive methods for multiple lung US classification tasks.
    Contrary to our initial belief, a contrastive method outperformed multiple non-contrastive methods on multiple lung US downstream tasks.
\end{itemize}
Figure~\ref{fig:overview} encapsulates the novel methods proposed in this study.
To the authors' knowledge, there are no preceding studies that systematically investigate the effect of sampling multiple images from the same US video in non-contrastive learning.
More generally, we believe that this study is the first to integrate sample weights into non-contrastive objectives.

\section{Background}
\label{sec:background}

\subsection{Joint Embedding Self-Supervised Learning}
\label{subsec:joint-embedding-ssl}
Having gained popularity in recent years in multiple imaging modalities, joint embedding SSL refers to a family of methods where the pretext task is to produce output vectors (i.e., \textit{embeddings}) that are close for examples satisfying a similarity pairwise relationship.
Pairs of images satisfying this relationship are known as \textit{positive pairs}, and they assumed to share semantic content with respect to the downstream task.
For example, positive pairs could belong to the same class in a downstream supervised learning classification task.
On the other hand, \textit{negative pairs} are pairs of images that do not satisfy the pairwise relationship.
In the label-free context of SSL, positive pairs are often constructed by sampling distorted versions of a single image~\cite{chen2020simple,grill2020bootstrap,zbontar2021barlow,bardes2022vicreg}.
The distortions are sampled from a distribution of sequentially applied transformations that are designed to preserve the semantic content of the image.
Horizontal reflection is a common example of a transformation that meets this criterion in many forms of imaging.

The architecture of joint embedding models commonly consists of two modules connected in series: a feature extractor and a projector.
The feature extractor is typically a convolutional neural network (CNN) or a variant of a vision transformer, while the projector is a multi-layer perceptron.
Following pre-training, the projector is discarded and the feature extractor is retained for weight initialization for the downstream task.

\textit{Contrastive learning} and \textit{non-contrastive learning} are two major categories of joint embedding methods.
Contrastive methods rely on objectives that explicitly attract positive pairs and repel negative pairs in embedding space.
Many of these methods adopt the InfoNCE objective~\cite{oh2016deep}, which may be viewed as cross-entropy for predicting which combination of embeddings in a batch correspond to a positive pair.
In most contrastive methods, positive pairs and negative pairs are distorted versions of the same image and different images, respectively.
MoCo is a contrastive method that computes pairs of embeddings using two feature extractors: a ``query" encoder and a ``key" encoder~\cite{he2020momentum}. 
The key encoder, which is an exponentially moving average of the query encoder, operates on negative examples
Its output embeddings are queued to avoid recomputation of negative embeddings.
SimCLR~\cite{chen2020simple} is a widely used contrastive method that employs a variant of the InfoNCE objective that does not include the embedding of the positive pair in the demoninator~\cite{oh2016deep}.
It does not queue negative embeddings, relying on large batches of negative examples.
NNCLR is a variation of SimCLR that assembles positive pairs from examples that are proximal to each other in embedding space~\cite{dwibedi2021little}.

Non-contrastive methods dispense with negative pairs altogether, limiting their focus to reducing the difference between embeddings of positive pairs.
They emerged as a way to ameliorate the large batch size that was previously thought to be necessary for contrastive methods.
It should be noted, however, that contrastive learning improvements have successfully reduced the batch size required to achieve state-of-the-art performance~\cite{yeh2022decoupled}.
By design, they address the information collapse problem -- a degenerate solution wherein all examples map to a null representation vector.
Self-distillation non-contrastive methods use architectural and asymmetrical training strategies to avoid collapse (e.g., BYOL~\cite{grill2020bootstrap}).
Information maximization non-contrastive methods address collapse by employing objectives that maximize the information content of the embedding dimensions.
For instance, the Barlow Twins method is a composite objective that contains a term for penalizing dimensional redundancy for batches of embeddings, in addition to a term for the distances between embeddings of individual positive pairs~\cite{zbontar2021barlow}.
VICReg introduced an additional term that explicitly maximizes variance across dimensions for batches of embeddings~\cite{bardes2022vicreg}.
Non-contrastive methods have been criticized for requiring embeddings with markedly greater dimensionality than the representations outputted by the feature extractor.

Recent theoretical works have sought to unify contrastive and non-contrastive methods.
Balestriero~\etal found that SimCLR, VICReg, and Barlow Twins are all manifestations of spectral embedding methods~\cite{Balestriero2022}.
Based on their results, they recommended that practitioners define a pairwise relationship that aligns with the downstream task.
For example, if the downstream task is classification, then positive pairs should have the same class.
Garrido~\etal challenged the widely held assumptions that non-contrastive methods perform better than contrastive methods and that non-contrastive methods rely on large embedding dimensions~\cite{garrido2022duality}.
They showed that the methods perform comparatively on benchmark tasks after hyperparameter tuning and that VICReg can be modified to reduce the dependence on large embeddings~\cite{garrido2022duality}.

\subsection{Joint Embedding Methods for B-Mode Lung Ultrasound}
\label{subsec:contrastive-methods-for-bmode-us}

Ultrasound is a dynamic imaging modality that is typically captured as a sequence of images and stored as a video.
As such, images originating from the same video are highly correlated and are likely to share semantic content.
Accordingly, recent works have developed US-specific contrastive learning methods that construct positive pairs from the same video.
The Ultrasound Contrastive Learning (USCL) method~\cite{chen2021uscl} is a derivative of SimCLR in which positive pairs are weighted sums of random images within the same video (i.e., the mixup operation~\cite{zhang2017mixup}), while negative pairs are images from different videos.
They reported an improvement on the downstream task of COVID-19 classification with the POCUS dataset~\cite{born2020pocovid}.
Improving on USCL, Meta-USCL concurrently trains a separate network that learns to weigh positive pairs~\cite{chen2022generating}.
The work was inspired by the observation that the intra-video positive pairs may exhibit a wide range of semantic similarity or dissimilarity.
Unfortunately, the training of the weighting network was biased, because its parameters were updated using the gradient of the objective calculated on the validation set. 
Basu~\etal proposed a MoCo-inspired solution where positive pairs are images that are temporally close within a video, while negative pairs consist of either pairs from different videos or pairs from the same video that are separated temporally by at least a threshold~\cite{basu_unsupervised_2022}.
Over the course of pretraining, the negative separation threshold is decreased, yielding increasingly difficult negative pairs.
Lastly, the HiCo method's objective is the sum of a softened InfoNCE loss calculated for the feature maps outputted by various model blocks~\cite{zhang2022hico}.
To soften InfoNCE, the one-hot positive/negative pair class labels were modified to slightly increase the entropy of the distribution.
The authors reported greatly improved performance with respect to USCL.

Standard non-contrastive methods have been applied for various tasks in US imaging.
In addition to assessing contrastive methods, Anand~\etal conducted pretraining with two self-distillation non-contrastive methods (BYOL~\cite{grill2020bootstrap} and DINO~\cite{caron2021emerging}) on a large dataset of echocardiograms~\cite{anand2022benchmarking}.
BYOL pretraining has also been applied successfully in anatomical tracking tasks on US~\cite{liang2023semi}.
Information maximization methods have been investigated for artifact detection tasks in M-mode and B-mode lung ultrasound~\cite{vanberlo2023exploring,vanberlo2023self}.
To our best knowledge, no studies have trialled non-contrastive learning methods for B-mode ultrasound with intra-video positive pairs.
The present study seeks to address this gap in the literature by investigating the effect of sampling positive pairs from the same video on the efficacy of non-contrastive pretraining for tasks in ultrasound.

\section{Methods}
\label{sec:methods}

\subsection{Joint Embedding Methods for Ultrasound with Intra-Video Positive Pairs}
\label{subsec:noncontrastive-us-pretraining}

\textbf{Setup:} We consider the standard joint embedding scenario where unlabelled data are provided and the goal is to maximize the similarity between embeddings of positive pairs.
In contrastive learning, the goal is augmented by maximizing the dissimilarity of negative pairs.
Let $\vb*{x}_1$ and $\vb*{x}_2$ denote a positive pair of US images.
Self-supervised pretraining results in a feature extractor $f(\vb*{x})$ that outputs representation vector $\vb*{h}$. The goal of SSL is to produce a feature extractor that is a better starting point for learning the downstream task than random initialization.

In this study, we propose a simple method for sampling and weighing positive pairs in the joint embedding setting that can be adopted for any joint embedding SSL method.
We adopt SimCLR~\cite{chen2020simple}, Barlow Twins~\cite{zbontar2021barlow} and VICReg~\cite{bardes2022vicreg} for our experiments.
In these methods, a MLP projector is appended to the feature extractor during pretraining.
$\vb*{z} = g(\vb*{h}) = g(f(\vb*{x}))$ is the embedding vector outputted by the projector.
The SSL objective is then computed in embedding space.

\textbf{Intra-video Positive Pairs: (IVPP)} Recall that positive pairs are images that are semantically related.
Previous work in contrastive SSL for US has explored the use of intra-video positive pairs~\cite{chen2021uscl,basu_unsupervised_2022,chen2022generating,zhang2022hico}.
A problem with naively sampling intra-video positive pairs is that it rests on the assumption that all videos in the same US video are sufficiently similar.
However, clinically relevant signs commonly surface and disappear within the same US video as the US probe and/or the patient move.
For example, B-lines are an artifact in lung US that signify diseased lung parenchyma.
B-lines may disappear and reappear as the patient breathes or as the sonographer moves the probe.
The A-line artifact appears in the absence of B-lines, indicating normal lung parenchyma.
In the absence of patient context, an image containing A-lines and an image containing B-lines from the same video convey very different impressions.
While most previous methods only considered inter-video images to be negative pairs, Basu~\etal recognized the problem of similar intra-video positive pairs by introducing hard negative sampling to their contrastive learning method -- temporally distant intra-video pairs of images were also considered negative pairs~\cite{basu_unsupervised_2022}.
However, there are situations in which distant intra-video images contain similar content.
For example, the patient and probe may remain stationary throughout the video, or the probe may return to its original position and/or orientation.
Without further knowledge of the US examinations in a dataset, we conjectured that it may be safest to only assume that positive pairs are intra-video images that are close to each other.
Closer pairs are likely to contain similar semantic content, yet they harbour different noise samples that models should be invariant to.

\begin{figure}
  \centering
  \begin{subfigure}{0.49\linewidth}
    \includegraphics[width=\linewidth]{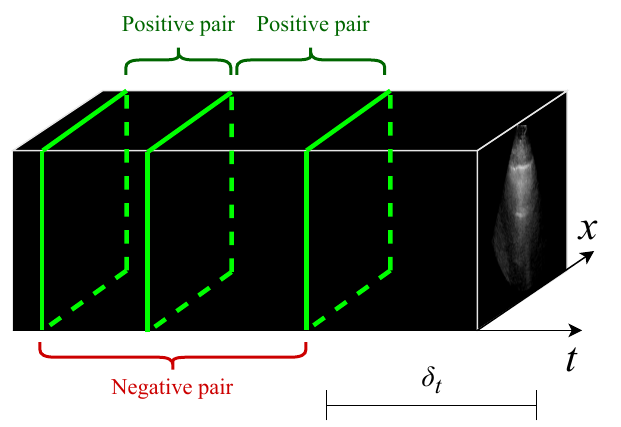}
    \vspace{10px}
    \caption{For B-mode ultrasound, positive pairs are frames in the same video that are within $\delta_t$ seconds of each other.}
    \label{subfig:bmode_pos_pairs}
  \end{subfigure}
  \hfill
  \begin{subfigure}{0.49\linewidth}
    \centering
    \includegraphics[width=0.7\linewidth]{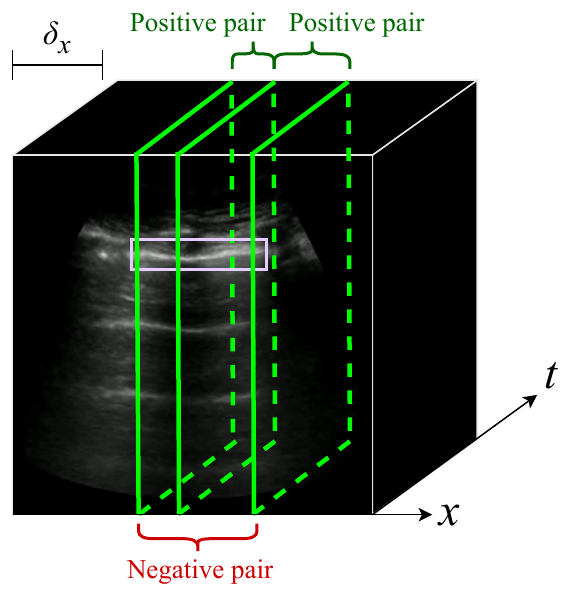}
    \caption{For M-mode ultrasound, positive pairs are M-mode images originating from the same B-video that are located within $\delta_x$ pixels from each other.
    In the context of lung ultrasound, M-mode images should intersect the pleural line (outlined in mauve).}
    \label{subfig:mmode_pos_pairs}
  \end{subfigure}
  \caption{
  Illustration of intra-video positive pairs. 
  Positive pairs are considered images that are no more than a threshold apart from each other within the same ultrasound video.
  }
  \label{fig:pos_pairs}
\end{figure}

For B-mode US videos, we define positive pairs as intra-video images $\vb*{x}_1$ and $\vb*{x}_2$ that are temporally separated by no more than $\delta_\text{max}$ seconds.
To accomplish this, $\vb*{x}_1$ is randomly drawn from the video's images, and $\vb*{x}_2$ is randomly drawn from the set of images that are within $\delta_\text{t}$ seconds of $\vb*{x}_1$.
Note that the frame rate of the videos must be known in order to determine which images are sufficiently close to $\vb*{x}_1$.

A similar sampling scheme is applied for M-mode US images.
Like previous studies, we define M-mode images as vertical slices through time of a B-mode video, taken at a specific x-coordinate in the video~\cite{VanBerlo2022,vanberlo2023exploring,javsvcur2021detecting}.
The x-axis of an M-mode image is time, and its y-axis is the vertical dimension of the B-mode video.
We define positive pairs to be M-mode images whose x-coordinates differ by no more than $\delta_\text{x}$ pixels.
To ensure spatial consistency, all B-mode videos are resized to the same width and height prior to sampling M-mode images.

As is customary in joint embedding methods, stochastic data augmentation is applied to each image, encouraging the feature extractor to become invariant to semantically insignificant differences.
Any data augmentation pipeline may be adopted for this formulation of intra-video positive pairs; however, we recommend careful selection of transformations and the distributions of their parameters to ensure that the pairwise relationship continues to be consistent with the downstream US task.

\textbf{Sample Weights: } The chance that intra-video images are semantically related decreases as temporal or spatial separation increases.
To temper the effect of unrelated positive pairs, we apply sample weights to positive pairs in the SSL objective according to their temporal or spatial distance.
Distant pairs are weighed less than closer pairs.
For a positive pair of B-mode images occurring at times $t_1$ and $t_2$ or M-mode images occurring at positions $x_1$ and $x_2$, the sample weight is calculated as follows:

\begin{align}
    w = \frac{\delta_t - |t_2 - t_1|}{\delta_t + 1} \qquad \qquad \qquad \qquad & w = \frac{\delta_x - |x_2 - x_1|}{\delta_x + 1}
    \label{eqn:sample-weight}
\end{align}

Sample weights are incorporated into each SSL objective.
The SimCLR objective~\cite{chen2020simple} can be easily modified by multiplying the per-example NT-Xent loss for the $i$\textsuperscript{th} positive pair $L_i$ by sample weight $w_i$.

\begin{equation}
    \mathcal{L}_\text{SimCLR} = \frac{1}{N} \sum_{i=1}^N w_i L_i
\end{equation}

For VICReg~\cite{bardes2022vicreg}, the invariance term is weighted with $w_{i} \in [0,1]$ for each positive pair in a batch. 
The invariance term is then calculated as follows:

\begin{equation}
    s(Z_1, Z_2) = \frac{1}{N} \sum_{i=1}^N w_i \|Z_{1_i} - Z_{2_j}\|_2^2
\end{equation}

where $Z_1$ and $Z_2$ are batches of predicted embeddings for corresponding positive pairs; that is, ${Z_1}_i$ and ${Z_2}_i$ correspond to one positive pair. 
The entire VICReg objective can then be calculated as

\begin{equation}
    \mathcal{L}_\text{VICReg}(Z_1, Z_2) = \underbrace{\lambda s(Z_1, Z_2)}_{\text{Invariance term}}
    + 
    \underbrace{\mu (v(Z_1) + v(Z_2))}_{\text{Variance term}}
    + 
    \underbrace{\nu(c(Z_1) + c(Z_2))}_{\text{Covariance term}}
    \label{eqn:weighted-vicreg}
\end{equation}

where $\lambda$, $\mu$, and $\nu$ are weights for each term.
Since frames are sampled uniformly at random, $\mathbb{E}[w] \simeq 0.5$.
Accordingly, we double $\lambda$ when pretraining VICReg with sample weights.

For the Barlow Twins objective, weighting is applied to each positive pair in the invariance term by computing the weighted normalized cross correlation matrix $\mathcal{C}_W \in \mathbb{R}^{D \times D}$ between the weighted-mean-centered normalized batches of embeddings, $Z_1$ and $Z_2$.
For a batch of embeddings $Z$, the calculation for the weighted mean $\Bar{Z}$ and standard deviations $\sigma(Z)$ across the batch dimension is performed as follows:

\begin{align}
    \bar{Z} = \frac{\sum_{i=1}^N { w_i} Z_i}
    {\sum_{i=1}^N { w_i} }    \qquad \qquad \qquad \qquad &
    \sigma({Z}) = \sqrt{
        \frac{\sum_{i=1}^N { w_i} (Z_i - \bar{Z})^2}
             {\sum_{i=1}^N { w_i} }
    }
    \label{eqn:w_moments}
\end{align}

where $w_i$ is the sample weight for the $i^\text{th}$ positive pair in the batch.
The redundancy reduction term should still be calculated using the normalized cross correlation matrix $\mathcal{C}$, since its purpose is to decorrelate the embedding dimensions.
In the original Barlow Twins, the normalized cross correlation matrix is employed for both terms.
The Barlow Twins objective then becomes

\begin{equation}
    \mathcal{L}_\text{BT} = 
    \underbrace{\sum_{\substack{d=1\\\vphantom{t}}}^D (1 - \mathcal{C}_{W_{d,d}})^2}_{\text{Invariance term}}
    \qquad +  
    \underbrace{\lambda \sum_{d=1}^D \sum_{\substack{e=1\\e \neq d}}^D{\mathcal{C}_{d,e}}^2}_{\text{Redundancy reduction term}}
    \label{eqn:weighted-barlow-twins}
\end{equation}

\subsection{Ultrasound Classification Tasks:}
\label{subsec:us-classification-tasks}

\textbf{COVID-19 Classification (\tt COVID): } As was done in previous studies on on US-specific joint embedding methods~\cite{chen2021uscl,basu_unsupervised_2022,chen2022generating,zhang2022hico}, we evaluate IVPP on the public POCUS lung US dataset~\cite{born2020pocovid}.
This dataset contains $140$ publicly sourced US videos ($2116$ images) labelled for three classes: COVID-19 pneumonia, non-COVID-19 pneumonia, and normal lung\footnote{See dataset details at the public POCUS repository~\cite{born2020pocovid}: \url{https://github.com/jannisborn/covid19_ultrasound}.}.
When evaluating on POCUS, we pretrain on the public Butterfly dataset, which contains $22$ unlabelled lung ultrasound videos~\cite{butte}\footnote{Accessed via a URL available at the public USCL repository~\cite{chen2021uscl}: \url{https://github.com/983632847/USCL}.}.

\textbf{A-line vs. B-line Classification ({\tt AB}): } A-lines and B-lines are two cardinal artifact in B-mode lung US that can provide quick information on the status of a patient's lung tissue.
A-lines are reverberation artifacts that are indicative of normal, clear lung parenchyma~\cite{soni2019point}. 
On lung US, they as horizontal lines deep to the pleural line.
Conversely, B-lines are indicative of diseased lung tissue~\cite{soni2019point}.
Generally, the two are mutually exclusive.
We evaluate on the binary classification task of A-lines versus B-lines on lung US, as was done in previous work benchmarking joint embedding SSL methods for lung US tasks~\cite{vanberlo2023self}.

We use a private dataset of \num{25917} parenchymal lung US videos (\num{5.9e6} images), hereafter referred to as \textit{ParenchymalLUS}.
ParenchymalLUS is a subset of a larger database of de-identified lung US videos that was partially labelled for previous work~\cite{Arntfield2021automation}.
Access to this database was permitted via ethical approval by Western University (REB 116838).
Before experimentation, we split the labelled portion of ParenchymalLUS by anonymous patient identifier into training, validation, and test sets.
The unlabelled portion of ParenchymalLUS was assembled by gathering \num{20000} videos from the unlabelled pool of videos in the database that were predicted to contain a parenchymal view of the lungs by a previously trained lung US view classifier~\cite{Vanberlo2022enhancing}.
All videos from the same patient were in either the labelled or the unlabelled subset.
Table~\ref{tab:dataset-description} provides further information on the membership of ParenchymalLUS.

\textbf{Lung Sliding Classification: ({\tt LS})} Lung sliding is a dymanic artifact that, when observed on a parenchymal lung US view, rules out the possibility of a pneumothorax at the site of the probe~\cite{Lichtenstein1995}.
The absence of lung sliding is suggestive of pneumothorax, warranting further investigation.
On B-mode US, lung sliding manifests as a shimmering of the pleural line~\cite{Lichtenstein1995}.
The presence or absence of lung sliding is also appreciable on M-mode lung US images that intersect the pleural line~\cite{Lichtenstein2005,Lichtenstein2010}.
We evaluate on the binary lung sliding classification task, where positive pairs are M-mode images originating from the same B-mode video.

ParenchymalLUS is adopted for the lung sliding classification task.
We use the same train/validation/test partition as described above.
Following prior studies, we estimate the horizontal bounds of the pleural line using a previously trained object detection model~\cite{VanBerlo2022} and use the top half of qualifying M-mode images, in decreasing order of total pixel intensity~\cite{vanberlo2023exploring}.

\begin{table*}[]
    \centering
    \setlength{\tabcolsep}{5pt}
    \begin{tabular}{cccccc}
    \toprule
    & & Unlabelled & \multicolumn{3}{c}{Labelled} \\
    & &  & Train & Validation & Test  \\
    \midrule
    \multirow{3}{*}{Total} & Patients &  $5204$ & $1540$ & $330$ & $329$  \\
    & Videos & \num{20000} & $4123$ & $858$ & $936$  \\
    & Images & \num{4611063} & \num{927889} & \num{191437} & \num{208648}  \\
    \midrule
    \multirow{2}{*}{A/B Line Labels} & Videos & $-$ & $2100\,/\,998$ & $441\,/\,197$ & $512\,/\,213$  \\
    & Images & $-$ & \num{484287}\,/\,\num{216505} & \num{99132}\,/\,\num{40608} & \num{116648}\,/\,\num{42122} \\
    \midrule
    \multirow{2}{*}{Lung Sliding Labels} & Videos & $-$ & $3169\,/\,477$ & $631\,/103$ & $707\,/\,96$  \\
    & Images & $-$ & \num{727205}\,/\,\num{96771} & \num{146322}\,/\,\num{23218} & \num{166753}\,/\,\num{21911}  \\
    \bottomrule
    \end{tabular}
    \caption{Training/validation/test breakdown of ParenchymalLUS for each task.
    $x\,/\,y$ indicates the number of negative and positive labelled examples available for each task.
    Video-level labels apply to all constituent images.
    Not all videos in the labelled }
    \label{tab:dataset-description}
\end{table*}

\section{Results}
\label{sec:results}

\subsection{Training Protocols}
\label{subsec:evaluation-protocol}

Unless otherwise stated, all feature extractors are initialized with ImageNet-pretrained weights.
Similar studies concentrating on medical imaging have observed that this practice improves downstream performance when compared to random initialization~\cite{azizi_big_2021,vanberlo2023exploring}.
Moreover, we designate fully supervised classifiers initialized with ImageNet-pretrained weights as a baseline against which to compare models pretrained with SSL.

Evaluation on POCUS follows a similar protocol employed in prior works~\cite{chen2021uscl,basu_unsupervised_2022}.
Feature extractors with the ResNet18 architecture~\cite{he2016deep} are pretrained on the Butterfly dataset. 
Prior to training on the POCUS dataset, a $3$-node fully connected layer with softmax activation was appended to the pretrained feature extractor.
Five-fold cross validation is conducted with POCUS by fine-tuning the final three layers of the pretrained feature extractor.
Unlike prior works, we adopt the average across-folds validation accuracy, instead of taking the accuracy of the combined set of validation set predictions across folds.

All experiments with ParenchymalLUS utilize the MobileNetV3-Small architecture as the feature extractor, which outputs a $576$-dimensional representation vector~\cite{howard2019searching}.
Feature extractors are pretrained on the union of the unlabelled videos and labelled training set videos in ParenchymalLUS.
Performance is assessed via test set classification metrics.
Prior to training on the downstream task, a single-node fully connected layer with sigmoid activation was appended to the pretrained feature extractor.
We report the performance of linear classifiers trained on the frozen feature extractor's representations, along with classifiers that are fine-tuned end-to-end.

For each joint embedding method, the projectors were multilayer perceptrons with two $768$-node layers, outputting $768$-dimensional embeddings.
Pretraining is conducted for $500$ epochs using the LARS optimizer~\cite{you2019large} with a batch size of $384$ and a learning rate schedule with warmup and cosine decay as in~\cite{bardes2022vicreg}.

The pretraining and training data augmentation pipelines consist of random transformations, including random cropping, horizontal reflection, brightness jitter, contrast jitter, and Gaussian blurring.
Additional data preprocessing details are available in Appendix~\ref{apx:pretraining-details}.

Source code will be made available upon publication\footnote{\url{https://github.com/bvanberl/IVPP}}.

\subsection{Performance}
\label{subsec:IVPP-performance}

The two main proposed features of IVPP are intra-video positive pairs and distance-based sample weights.
Accordingly, we assess the performance of IVPP across multiple assignments of the maximum image separation.
Separate trials were conducted for SimCLR, Barlow Twins, and VICReg pretraining.
For the {\tt COVID} and {\tt AB} tasks, we explored the values $\delta_t \in \{0, 0.5, 1, 1.5\}$ seconds.
The {\tt LS} task is defined for M-mode US, and so we explored $\delta_x \in \{0, 5, 10, 15\}$ pixels.
The standardized width of B-mode US videos should be considered when determining an appropriate range for $\delta_x$.
Note that when $\delta = 0$, sample weights are all $1$ and therefore do not modify any of the SSL objectives investigated in this study. 

\begin{figure}[]
    \centering
    \includegraphics[width=0.5\textwidth]{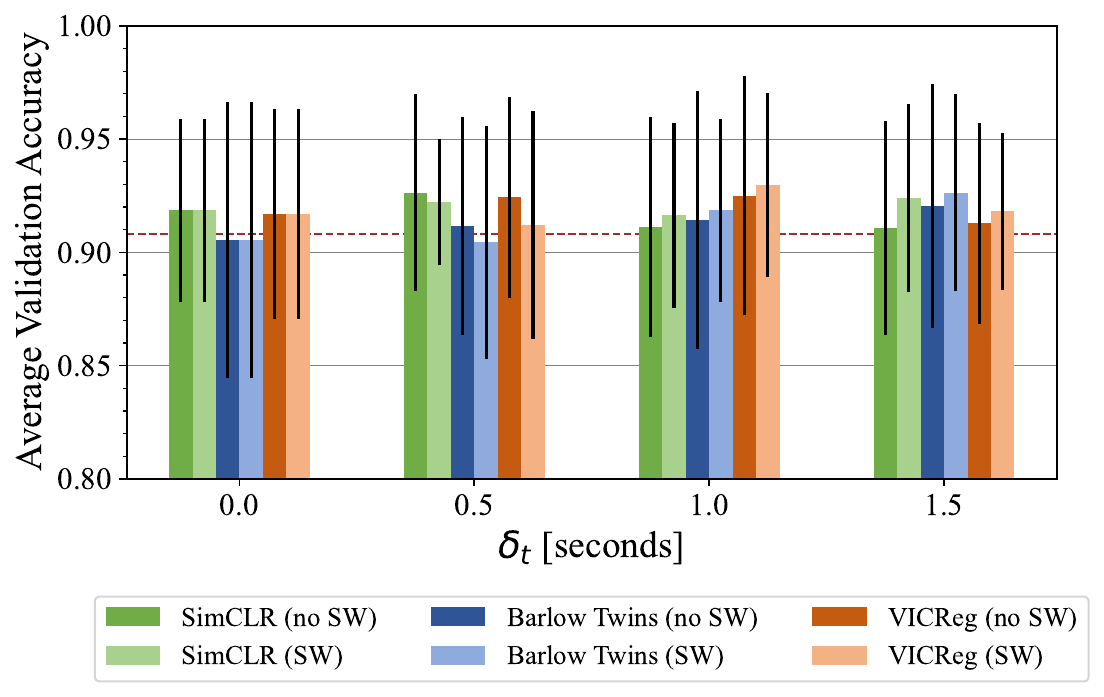}
    \caption{
    Average test accuracy across $5$-fold cross validation on the POCUS dataset. 
    Models were pretrained with a variety of intra-video positive pair thresholds with and without sample weights. 
    Error bars indicate the standard deviation across folds.
    The dashed line indicates initialization with ImageNet-pretrained weights.
    }
    \label{fig:POCUS_cv_results}
\end{figure}

Figure~\ref{fig:POCUS_cv_results} summarizes the performance of IVPP on the public POCUS dataset after pretraining on the Butterfly dataset, which is measured by average test accuracy in $5$-fold cross validation. 
In most cases, pretrained models obtained equal or greater average accuracy than the ImageNet-pretrained baseline, with the exception of Barlow Twins with $\delta_t = 0$ and $\delta_t = 0.5$.
The performance of models pretrained with SimCLR, Barlow Twins, and VICReg peaked at different nonzero values of $\delta_t$ ($0.5$, $1$, and $1.5$ respectively), indicating a possible benefit of selecting temporally close yet distinct intra-video positive pairs.
It was also observed across all three pretraining methods that the inclusion of sample weights resulted in worsened test AUC when $\delta = 0.5$, but improved test AUC when $\delta = 1.0$ and $\delta = 1.5$.

\begin{figure}
  \centering
  \begin{subfigure}{0.49\linewidth}
    \includegraphics[width=\linewidth]{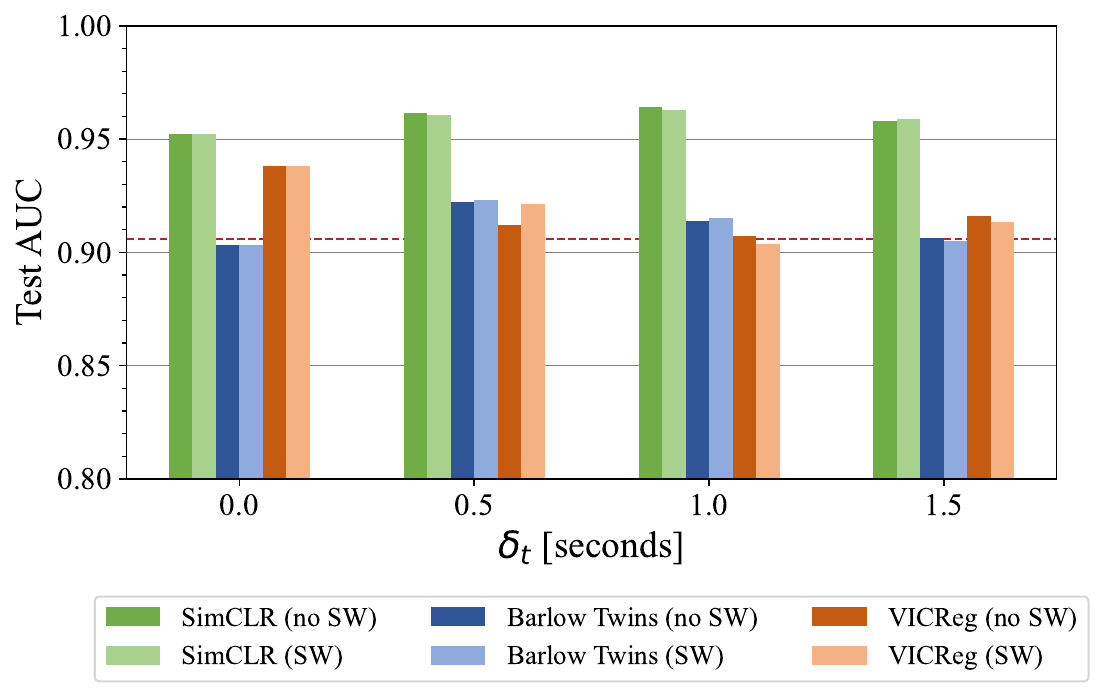}
    \caption{Linear classifiers for the {\tt AB} task}
    \label{subfig:linear-ab}
  \end{subfigure}
  \hfill
  \begin{subfigure}{0.49\linewidth}
    \includegraphics[width=\linewidth]{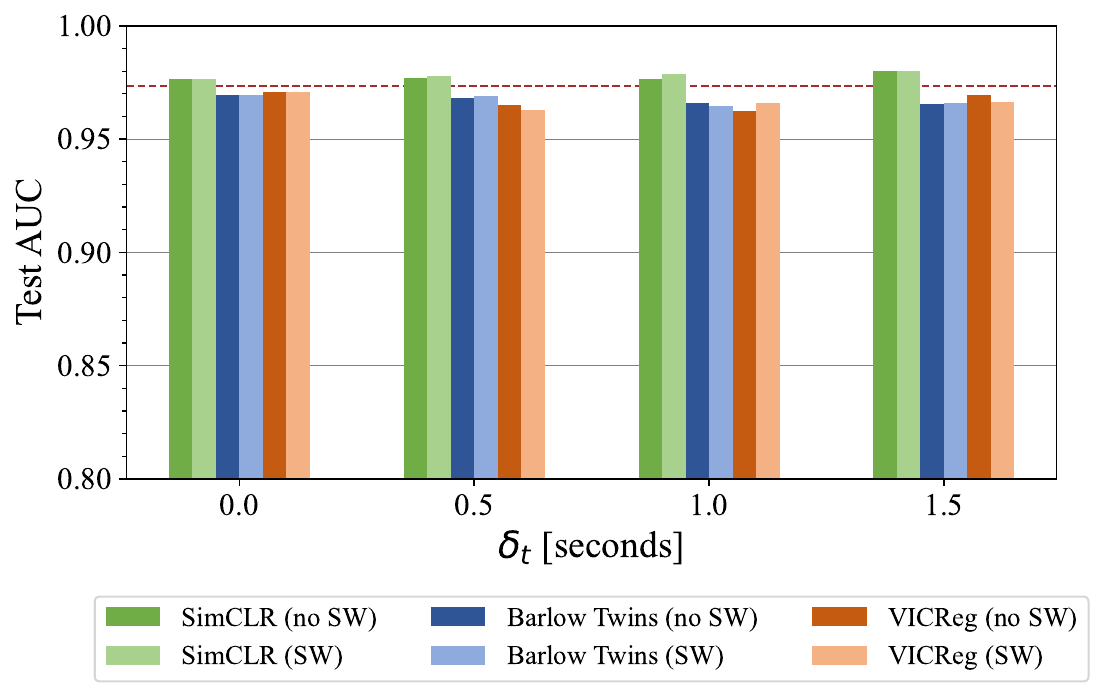}
    \caption{Fine-tuned classifiers for the {\tt AB} task}
    \label{subfig:finetune-ab}
  \end{subfigure}
  \begin{subfigure}{0.49\linewidth}
    \includegraphics[width=\linewidth]{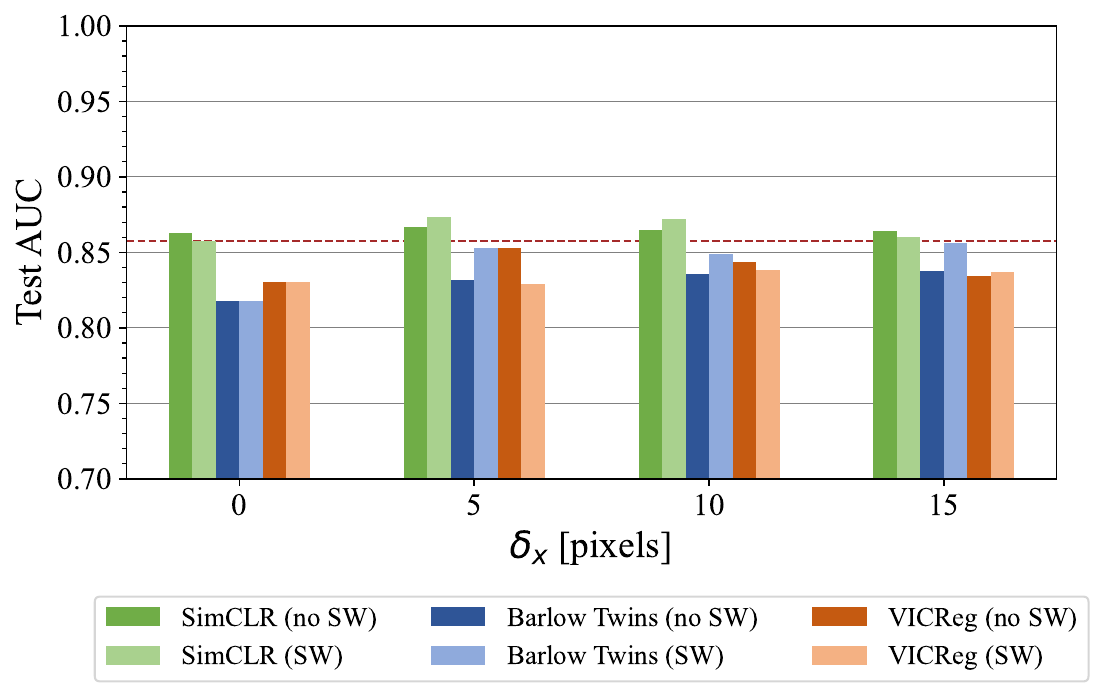}
    \caption{Linear classifiers for the {\tt LS} task}
    \label{subfig:linear-ls}
  \end{subfigure}
  \hfill
  \begin{subfigure}{0.49\linewidth}
    \includegraphics[width=\linewidth]{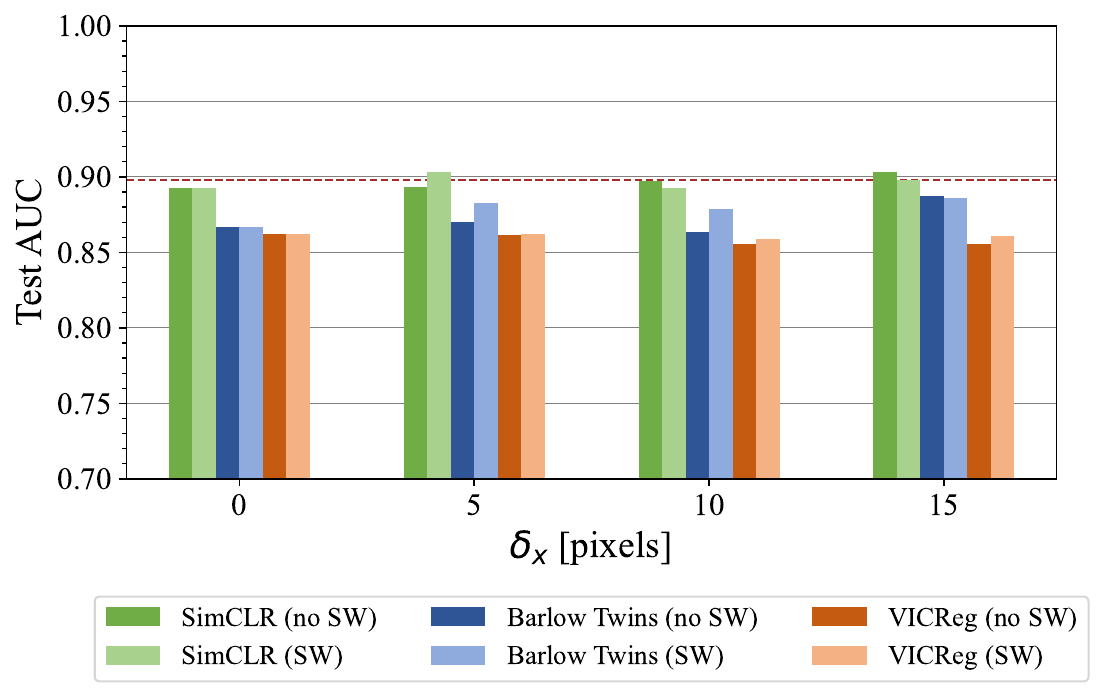}
    \caption{Fine-tuned classifiers for the {\tt LS} task}
    \label{subfig:finetune-ls}
  \end{subfigure}
  \caption{
  ParenchymalLUS test set AUC for the {\tt AB} and {\tt LS} binary classification tasks, calculated for models pretrained with a selection of contrastive and non-contrastive learning methods and employing a variety of intra-video positive pair thresholds with and without sample weights (SW). 
  The dashed line indicates initialization with ImageNet-pretrained weights.
  }
  \label{fig:parenchymallus-test-performance}
\end{figure}

Similar experiments were conducted with ParenchymalLUS for the {\tt AB} task and {\tt LS} task, using B-mode and M-mode images respectively as input.
ParenchymalLUS represents a scenario where there is an abundance of unlabelled data, which differs greatly from the preceding evaluation on public, yet small, datasets.
The unlabelled and labelled portions of ParenchymalLUS contained at least an order of magnitude more videos than either the public Butterfly and POCUS datasets.
B-mode and M-mode feature extractors were pretrained on the union of the unlabelled and training portions of ParenchymalLUS -- one for each value of $\delta$, with and without sample weights.
For these evaluations, we use all training examples that have been assigned a label for the downstream task.
Figure~\ref{fig:parenchymallus-test-performance} provides a visual comparison of the test AUC obtained by linear feature representation classifiers and fine-tuned models for the {\tt AB} and {\tt LS} tasks.
An immediate trend across both tasks and evaluation types is that SimCLR consistently outperformed Barlow Twins and VICReg, which are both non-contrastive methods.
Furthermore, pretraining with non-contrastive methods often resulted in worse test AUC compared to initialization with ImageNet-pretrained weights.
Another observation across all experiments was that there was no discernible trend for the effect of sample weights that was consistent for any task, pretraining method, $\delta_t$, or $\delta_x$.

Focusing on {\tt AB}, linear classifiers achieved the greatest performance when $\delta_t > 0$, with the exception of VICReg (Figure~\ref{subfig:linear-ab}).
The use of SimCLR compared to the other pretraining methods appeared to be responsible for the greatest difference in test performance.
As shown in Figure~\ref{subfig:linear-ab}, SimCLR-pretrained models outperformed non-contrastive methods, and were the only models that outperformed ImageNet-pretrained weights.
The use of a nonzero $\delta_t$ resulted in slight improvement in combination with SimCLR pretraining, but degraded performance of non-contrastive methods.

Similar results were observed for the {\tt LS} M-mode classification task.
Models pretrained with SimCLR were the only ones that matched or surpassed fully supervised models.
Nonzero $\delta_x$ generally improved the performance of linear classifiers, with $\delta_x = 5$ pixels corresponding to the greatest test AUC for SimCLR and VICReg, and $\delta_x = 15$ for Barlow Twins.
Inclusion of sample weights appreciably improved the performance of Barlow Twins-pretrained models.
Fine-tuned models pretrained with SimCLR performed similarly to fully supervised models, while non-contrastive methods resulted in degradation of test AUC.

Table~\ref{tab:performance} compares the top-performing IVPP-pretrained models for each SSL method with two prior US-specific contrastive learning methods -- USCL~\cite{chen2021uscl} and US~UCL~\cite{basu_unsupervised_2022}.
Of note is that all three self-supervised methods pretrained with IVPP outperformed ImageNet-pretrained initialization for POCUS, a task where very little pretraining \textit{and} training data were utilized.
For the B-mode and M-mode tasks assessed with ParenchymalLUS, contrastive methods (including the baseline) outperformed non-contrastive methods.
Overall, the most pronounced result apparent from the above experiments is that SimCLR, a contrastive method, outperformed both non-contrastive methods when unlabelled data is abundant.



\begin{table}
    \centering
    \begin{tabular}{lccc}
        \toprule
        Dataset & POCUS & \multicolumn{2}{c}{ParenchymalLUS}  \\[5pt]
        Pretraining method & Mean (std) test accuracy & A/B Test AUC & LS Test AUC \\
        \midrule
        Random initialization & $0.881\,(0.050)$ & $0.954$ & $0.790$ \\
        ImageNet initialization & $0.908\, (0.043)$ & $0.973$ & $0.898$ \\
        USCL~\cite{chen2021uscl} & $0.905\, (0.044)$ & $0.979$ & $0.874$ \\
        UCL~\cite{basu_unsupervised_2022} & $0.901\, (0.054)$ & $0.967$ & $0.809$ \\
        IVPP (SimCLR) & $0.926\, (0.043)$ & $0.980$ & $0.903$ \\
        IVPP (Barlow Twins) & $0.921\, (0.054)$ & $0.969$ & $0.887$ \\
        IVPP (VICReg) & $0.930\, (0.046)$ & $0.971$ & $0.862$ \\
        \bottomrule
    \end{tabular}
    \caption{Performance of fine-tuned models pretrained using IVPP compared to US-specific contrastive learning methods, USCL and UCL, and to baseline Random and ImageNet initializations.}
    \label{tab:performance}
\end{table}

\subsection{Label Efficiency}

ParenchymalLUS is much larger than public ultrasound datasets for machine learning.
Although the majority of its videos are unlabelled, it contains a large number of labelled examples.
To simulate a scenario where the fraction of examples that are labelled is much smaller, we investigated the downstream performance of models that were pretrained on all the unlabelled and training ParenchymalLUS examples and then fine-tuned on a very small subset of the training set.

Conventionally, label efficiency investigations are conducted by fitting a model for the downstream task using progressively smaller fractions of training data to gauge how well self-supervised models fare in low-label scenarios.
The results of these experiments are likely influenced by the particular training subset that is randomly selected.
We designed an experiment to determine if the choice of $\delta_t$, $\delta_x$, or the introduction of sample weights influenced downstream performance in low-label settings.
To reduce the chance of biased training subset sampling, we divided the training set into $20$ subsets and repeatedly performed fine-tuning experiments on each subset for each pretraining method and $\delta$ value, with and without sample weights.
For each pretraining method, two-way repeated-measures analysis of variance (ANOVA) was performed to determine whether the mean test AUCs across values of $\delta$ and sample weight usage were different.
The independent variables were $\delta$ and the presence of sample weights, while the dependent variable was test AUC.
Whenever the null hypothesis of the ANOVA was rejected, post-hoc paired \textit{t}-tests were performed to compare the following:
\begin{itemize}
    \item Pretraining with nonzero $\delta$ against standard positive pair selection ($\delta=0$)
    \item For the same nonzero $\delta$ value, sample weights against no sample weights
\end{itemize}
For each group of post-hoc tests, the Bonferroni correction was applied to establish a family-wise error rate of $\alpha = 0.05$.
To ensure that each training subset was independent, we split the dataset by anonymous patient identifier.
This was a necessary step because intra-video images are highly correlated, along with videos from the same patient.
As a result, the task became substantially more difficult than naively sampling $5\%$ of training images because the volume \textit{and} heterogeneity of training examples was reduced by training on a small fraction of examples from a small set of patients.

\begin{figure}
  \centering
  \begin{subfigure}{0.49\linewidth}
    \includegraphics[width=\linewidth]{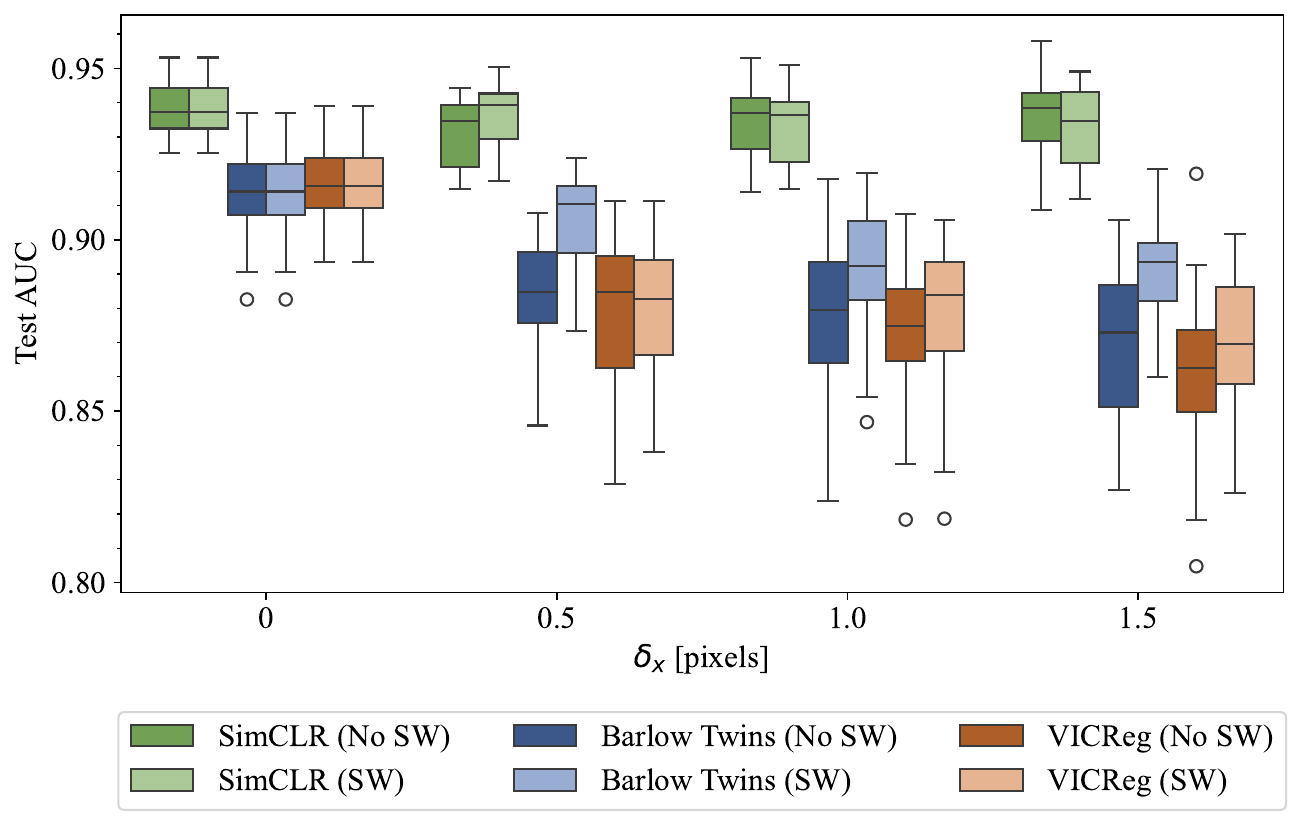}
    \caption{{\tt AB} task}
    \label{subfig:boxplot-ab}
  \end{subfigure}
  \hfill
  \begin{subfigure}{0.49\linewidth}
    \includegraphics[width=\linewidth]{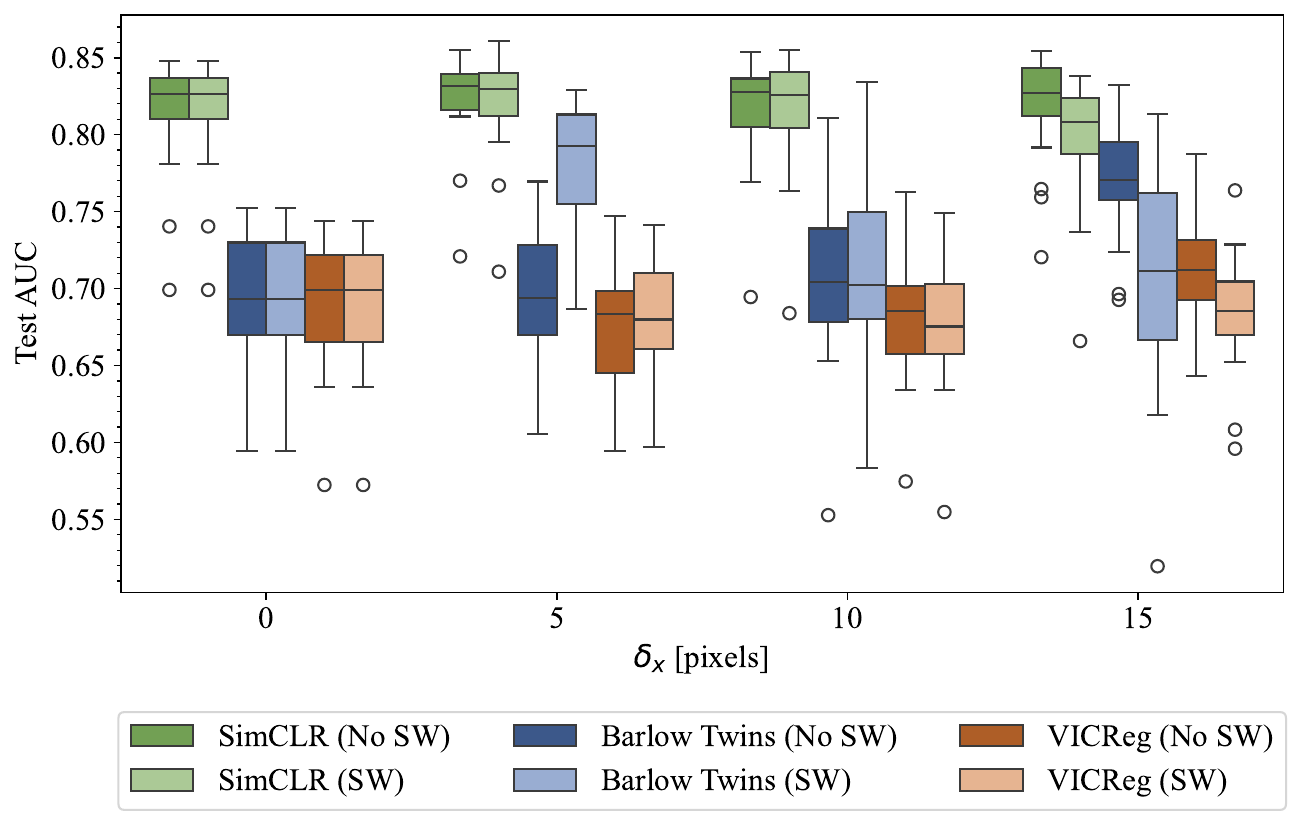}
    \caption{{\tt LS} task}
    \label{subfig:boxplot-ls}
  \end{subfigure}
  \caption{
  Boxplots conveying the quartile ranges of test AUC distributions for each pretraining method and assignment to $\delta$, with and without sample weights.
  Each experiment is repeated $20$ times on disjoint subsets of the training set, each containing all images from a group of patients.
  }
  \label{fig:boxplots}
\end{figure}

The fine-tuning procedure was identical to that described in section~\ref{subsec:evaluation-protocol}, with the exception that the model's weights at the end of training were retained for evaluation, instead of restoring the best-performing weights on the validation set.
Figure~\ref{fig:boxplots} provides boxplots for all trials that indicate the distributions of test AUC under the varying conditions for both the {\tt AB} and {\tt LS} tasks.
Again, SimCLR performance appeared to be substantially higher than both non-contrastive methods.

Table~\ref{tab:pairwise-t-tests} gives the mean and standard deviation of each set of trials, for each hyperparameter combination.
For each task and each pretraining method, the ANOVA revealed significant interaction effects ($p \le 0.05$).
Accordingly, all intended post-hoc \textit{t}-tests were performed to ascertain (1) which combinations of hyperparameters were different from the baseline setting of augmenting the same frame twice ($\delta = 0$) and (2) values of $\delta$ where the addition of sample weights changes the outcome.
First, we note that SimCLR was the only pretraining method that consistently outperformed full supervision with ImageNet-pretrained weights.
Barlow Twins and VICReg pretraining -- both non-contrastive methods -- resulted in worse performance.

For the {\tt AB} task, no combination of intra-video positive pairs or sample weights resulted in statistically significant improvements compared to dual distortion of the same image ($\delta_t = 0$). 
For Barlow Twins and VICReg, several nonzero $\delta_t$ resulted in significantly worse mean test AUC.
Sample weights consistently made a difference in Barlow Twins across $\delta_t$ values, but only improved mean test AUC for $\delta_t=1$ and $\delta_t=1.5$.

Different trends were observed for the {\tt LS} task. 
SimCLR with $\delta_x = 5$ and no sample weights improved mean test AUC compared to the baseline where $\delta_x = 0$.
No other combination of hyperparameters resulted in a significant improvement.
For Barlow Twins, multiple IVPP hyparameter combinations resulted in improved mean test AUC over the baseline.
No IVPP hyperparameter combinations significantly improved the performance of VICReg.

\begin{table}[h]
    \begin{threeparttable}
    \centering
    \begin{tabular}{c@{\hskip 0.6in}ccc@{\hskip 0.6in}ccc}
        \toprule
        & \multicolumn{3}{c}{{\large \tt AB}} & \multicolumn{3}{c}{{\large \tt LS}} \\[5pt]
         Pretraining Method & $\delta_t$ & SW & Mean (std) test AUC & $\delta_x$ & SW & Mean (std) test AUC \\
         \midrule
         \multirow{7}{*}{{\large SimCLR}} & $0$ & \xmark & $0.938\,(0.007)$ & $0$ & \xmark & $0.812\,(0.037)$  \\
         & $0.5$ & \xmark & $0.931\,(0.010)\,^*$ & $5$ & \xmark &  $0.824\,(0.030)\,^{*}$ \\
         & $0.5$ & \cmark & $0.936\,(0.007)\,^\dagger$ & $5$ & \cmark & $0.820\,(0.033)$ \\
         & $1$ & \xmark & $0.934\,(0.011)$ & $10$ & \xmark & $0.815\,(0.035)$ \\
         & $1$ & \cmark & $0.933\,(0.011)$ & $10$ & \cmark & $0.816\,(0.037)$ \\
         & $1.5$ & \xmark & $0.936\,(0.013)$ & $15$ & \xmark & $0.819\,(0.034)$ \\
         & $1.5$ & \cmark & $0.932\,(0.012)$ & $15$ & \cmark & $0.798\,(0.039)\,^{*\dagger}$ \\
         \midrule
         \multirow{7}{*}{{\large Barlow Twins}} & $0$ & \xmark & $0.914\,(0.014)$ & $0$ & \xmark & $0.693\,(0.044)$ \\
         & $0.5$ & \xmark & $0.914\,(0.010)\,^*$ & $5$ & \xmark &  $0.694\,(0.040)$ \\
         & $0.5$ & \cmark & $0.883\,(0.017)\,^{*\dagger}$ & $5$ & \cmark & $0.780\,(0.040)\,^{*\dagger}$ \\
         & $1$ & \xmark & $0.877\,(0.022)\,^*$ & $10$ & \xmark & $0.705\,(0.051)$ \\
         & $1$ & \cmark & $0.891\,(0.018)\,^{*\dagger}$ & $10$ & \cmark & $0.706\,(0.066)$ \\
         & $1.5$ & \xmark & $0.870\,(0.024)\,^*$ & $15$ & \xmark & $0.769\,(0.037)\,^{*}$ \\
         & $1.5$ & \cmark & $0.892\,(0.015)\,^{*\dagger}$ & $15$ & \cmark & $0.707\,(0.071)\,^{\dagger}$ \\
         \midrule
         \multirow{7}{*}{{\large VICReg}} & $0$ & \xmark & $0.917\,(0.011)$ & $0$ & \xmark & $0.690\,(0.042)$ \\
         & $0.5$ & \xmark & $0.879\,(0.024)\,^{*}$ & $5$ & \xmark &  $0.675\,(0.036)$ \\
         & $0.5$ & \cmark & $0.879\,(0.021)\,^{*}$ & $5$ & \cmark & $0.679\,(0.038)$ \\
         & $1$ & \xmark & $0.872\,(0.023)\,^{*}$ & $10$ & \xmark & $0.680\,(0.039)$ \\
         & $1$ & \cmark & $0.876\,(0.024)\,^{*}$ & $10$ & \cmark & $0.675\,(0.040)$ \\
         & $1.5$ & \xmark & $0.860\,(0.026)\,^{*}$ & $15$ & \xmark & $0.710\,(0.036)$ \\
         & $1.5$ & \cmark & $0.870\,(0.021)\,^{*\dagger}$ & $15$ & \cmark & $0.685\,(0.039)\,^{\dagger}$ \\
         \midrule
        None (ImageNet-pretrained) & & & $0.896\,(0.017)$ & & & $0.783\,(0.028)$ \\
        None (random initialization) & & & $0.774\,(0.051)$ & & & $0.507\,(0.022)$ \\
        \bottomrule
    \end{tabular}
    \begin{tablenotes}
    \item[$*$] Significantly different ($p < 0.05$) than baseline for the pretraining method where $\delta = 0$ 
    \item[$\dagger$] Significantly different ($p < 0.05$) for particular $\delta$ when sample weights are applied, compared to no sample weight 
    \end{tablenotes}
    \caption{
    ParenchymalLUS test AUC for the the {\tt AB} and {\tt LS} tasks when trained using examples from $5\%$ of the patients in the training set. 
    Twenty trials were performed for each pretraining method, value of $\delta$, with and without sample weights (SW).
    Mean and standard deviation of the test AUC across trials are reported for each condition.
    }
    \label{tab:pairwise-t-tests}
    \end{threeparttable}
\end{table}

\section{Discussion}
\label{sec:discussion}

\subsection{Guidelines for Practitioners}
Multiple insights can be derived from the results that will guide practitioners working with deep learning for ultrasound interpretation.
Firstly, SimCLR was observed to achieve the greatest performance consistently across multiple tasks.
With the exception of the data-scarce COVID-19 classification task, SimCLR decisively outperformed Barlow Twins and VICReg.
Surprisingly, non-contrastive methods often exhibited inferior performance compared to full supervision with ImageNet-pretrained weight initialization.
The results provide evidence towards favouring contrastive methods over non-contrastive methods for problems in ultrasound.
It could be that the non-contrastive methods studied may not be suited toward the US modality.
These observations contradict theoretical results claiming that downstream error is more likely to be reduced with VICReg and Barlow Twins than with SimCLR when positive pair alignment is not guaranteed~\cite{balestriero2017neural}.
Future work assessing non-contrastive methods for tasks in different ultrasound examinations or alternative imaging modalities altogether would shed light on the utility of non-contrastive methods outside the typical evaluation setting of photographic images.

While the experimental results do not support the existence of overarching trends for hyperparameter assignments for intra-video positive pairs across pretraining methods, it was observed that some combinations improved performance on downstream tasks. 
For the COVID-19 classification task, each pretraining method was improved by a nonzero value of $\delta_t$.
Sample weights improved performance for larger values of $\delta_t$, but worsened performance for smaller $\delta_t$.
For the {\tt AB} task, there existed at least one nonzero $\delta_t$ that improved performance for each pretraining method, when fine-tuning on all available training data.
Sample weights appeared to offer a slight boost to SimCLR-pretrained models for the top-performing values of $\delta_t$.
The same was true for the {\tt LS} task, but only when SimCLR was the pretraining method.
Interestingly, intra-video positive pairs did not provide any benefit to SimCLR when training {\tt AB} classifiers on small fractions of patients represented in the training set.
For the {\tt LS} task, a statistically significant performance increase was observed when using intra-video positive pairs with little separation ($\delta_x=5$, no sample weights).
Overall, the results indicate that the optimal assignment for IVPP hyperparameters may be problem-specific.
Intra-video positive pairs may improve performance on downstream ultrasound interpretation tasks; however, practitioners are advised to include a range of values of $\delta$ with and without sample weights in their hyperparameter search.

\subsection{Limitations}

The methods and experiments conducted in this study were not without limitations.
As is common in medical imaging datasets, the ParenchymalLUS dataset was imbalanced.
The image-wise representation for the positive class was $30.0\%$ for the {\tt AB} task and $11.7\%$ for the lung sliding task.
Although some evidence exists in support for self-supervised pretraining for alleviating the ill effects of class imbalance in photographic images~\cite{yang2020rethinking,liu2021self}, computed tomography, and fundoscopy images~\cite{zhang2023dive}, there is no such evidence for tasks in medical ultrasound.

As outlined in the background, the pretraining objectives employed in this study have been shown to improve downstream performance when the pairwise relationship aligns with the downstream task~\cite{Balestriero2022}.
These guarantees compare to the baseline case of random weight initialization.
While it was observed that all pretraining methods outperformed full supervision with randomly initialized weights, ImageNet-pretrained weights outperformed non-contrastive methods in several of the experiments.
ImageNet-pretrained weights are a strong and meaningful baseline for medical imaging tasks, as they have been shown to boost performance in several supervised learning tasks across medical imaging modalities~\cite{azizi_big_2021}.
It is possible that some extreme data augmentation transformations and intra-positive pairs could jeopardize the class agreement of positive pairs (as is likely in most pragmatic cases); however, near-consistent alignment was achieved through data augmentation design and small ranges of $\delta$.
Although there exists evidence that VICReg and SimCLR can achieve similar performance on ImageNet with judicious selection of hyperparameters (e.g., temperature, loss term weights, learning rate)~\cite{garrido2022duality}, experiments in this study utilized default hyperparameters, the explosion of hyperparameter space when combined with those explored in this study in the context of limited computational resources.

Lastly, M-mode images were designated by selecting $x$-coordinates in B-mode videos that intersect a pleural line region of interest, as predicted by an object detection model utilized in previous work~\cite{VanBerlo2022,vanberlo2023exploring}.
LUS M-mode images must intersect the pleural line in order to appreciate the lung sliding artifact.
While we mitigate potential inaccuracies in localization by limiting training and evaluation data to the brightest half of eligible $x$-coordinates, it is possible that a small fraction of M-mode images were utilized that did not intersect the pleural line.


\subsection{Future Work}

There are multiple avenues for further investigation that could complement and extend the findings of this study.
First, IVPP could be studied for a variety of non-lung medical ultrasound exams, such as echocardiography (B-mode or M-mode) or thyroid ultrasound (B-mode).
On a related note, the dominance of SimCLR on the downstream tasks of A-line versus B-line classification and lung sliding detection warrant more investigations comparing contrastive to non-contrastive methods for different tasks in ultrasound.
IVPP could also be integrated into other SSL objectives, such as BYOL~\cite{grill2020bootstrap}.
The sample weights formulation proposed in this study could also be applied to SSL for non-ultrasound videos.
For example, VITO is a contrastive learning method for images that uses intra-video positive pairs~\cite{parthasarathy2023selfsupervised}.
Future work should also compare contrastive and non-contrastive methods for other SSL methods and types of ultrasound examinations, which would support or refute this paper's conclusion about the superiority of contrastive methods for ultrasound images.
Lastly, future work concentrating on SSL in ultrasound could also focus on discovering ultrasound-specific data augmentation transformations that preserve semantic content, and investigate their effect on downstream performance when compared to the standard transformations relied on by joint embedding methods.
As a natural source of differences between positive pairs, IVPP could be studied in tandem with the use of such augmentations.

\section{Conclusion}

Intra-video positive pairs have been proposed as a means of improving the downsteam performance of ultrasound classifiers pretrained with joint embedding self supervised learning.
In this study, we suggested a scheme for integrating such positive pairs into common contrastive and non-contrastive SSL methods.
Applicable to both B-mode and M-mode ultrasound, the proposed method (IVPP) consists of sampling positive pairs that are separated temporally or spatially by no more than a threshold, optionally applying sample weights to each pair in the objective depending on the distance.
A thorough investigation of IVPP revealed that using nearby images from the same video for positive pairs can lead to improved performance when compared to composing positive pairs from the same image, but that IVPP hyperparameter assignments yielding benefits may vary by the downstream task.
Another salient result was the persistent top performance of SimCLR for key tasks in B-mode and M-mode lung ultrasound, indicating that contrastive learning may be more suitable than non-contrastive learning methods for ultrasound imaging.
Subsequent work should continue to search for ways to take advantage of the uniquely videographic nature of ultrasound.

\section*{Acknowledgements}

B. VanBerlo is a Vanier Scholar (FRN 186945) supported by the Natural Sciences and Engineering Research Council of Canada (NSERC).
Computational resource support was also provided by Compute Ontario (\href{https://www.computeontario.ca}{computeontario.ca} and the Digital Research Alliance of Canada (\href{https://alliancecan.ca}{alliance.can.ca}).

\section*{Appendix}

\begin{appendix}

\section{Image Preprocessing}
\label{apx:pretraining-details}

B-mode and M-mode images were resized to $224\times224$ and $224\times112$ pixels respectively using bilinear interpolation.
The reason that the width of M-modes was standardized to a smaller value was because the height of B-mode images often far exceeded the number of frames in a $3$-second segment of B-mode video.
Since feature extractors were initialized with pretrained weights, pixel intensities were mean-centered and normalized using the mean and standard deviation of the ImageNet dataset.

All IVPP pretraining runs were subjected to stochastic data augmentations after intra-video positive pairs were sampled.
Each image was subjected to the following sequence of stochastic transformations for data augmentation:
\begin{enumerate}
    \item Randomly located crop of a fraction of the image's area in the range $[0.4, 1.0]$, followed by resizing to the original image dimensions. For B-mode images, the width/height aspect ratio was confined to the range $[0.8, 1.25]$, while M-mode crops were confined to  $[0.4, 0.6]$.
    \item With probability $0.5$, horizontal reflection.
    \item With probability $0.5$, random brightness change in the range $[-0.25, 0.25]$.
    \item With probability $0.5$, random contrast change in the range $[-0.25, 0.25]$.
    \item With probability $0.25$, random Gaussian blur with a kernel size of $5$ pixels and a standard deviation uniformly sampled from the range $[0.1, 2.0]$.
\end{enumerate}
Training runs for the {\tt COVID} and {\tt AB} tasks with B-mode images also utilized this data augmentation pipeline.
For {\tt LS} training runs with M-modes, the minimum allowable crop area was increased to $95\%$ of the image's area to ensure the pleural line was almost always visible.

\section{Detailed Results}

Results are reported for all of the experiments described in Section~\ref{subsec:IVPP-performance}.
Average test accuracy for all values of $\delta_t$ with and without sample weights are given in Table~\ref{tab:pocus-results}.
Table~\ref{tab:linear-results} details the performance of linear classifiers trained on representations outputted by a pretrained feature extractor.
Lastly, Table~\ref{tab:fine-tune-results} provides the results of fine-tuning pretrained feature extractors. 
The metrics provided in these tables are the same as those conveyed in Figures~\ref{fig:POCUS_cv_results}~and~\ref{fig:parenchymallus-test-performance} as bar charts.

\section{Experimental Tools}

All pretraining and training runs were conducted using virtual machines equipped with an Intel E5-2683 v4 Broadwell CPU at \SI{2.1}{GHz} and a Nvidia Tesla P100 GPU with \SI{12}{GB} of VRAM.
Python 3.10 was utilized for all IVPP experiments.
The official USCL\footnote{\url{https://github.com/983632847/USCL}} and UCL\footnote{\url{https://github.com/gbc-iitd/US_UCL}} source code repositories were adapted for executing experiments comparing them to IVPP.

\begin{table}[h!]
    \small
    \centering
    \begin{tabular}{c@{\hskip 0.6in}ccc}
        \toprule
         Pretraining Method & $\delta_t$ & SW & Mean (std) Test Accuracy \\
         \midrule
         \multirow{7}{*}{{\large SimCLR}} & $0$ & \xmark & $0.919\, (0.041)$  \\
         & $0.5$ & \xmark & $0.926\,(0.043)$  \\
         & $0.5$ & \cmark & $0.922\,(0.028)$ \\
         & $1$ & \xmark & $0.911\,(0.049)$ \\
         & $1$ & \cmark & $0.916\,(0.041)$  \\
         & $1.5$ & \xmark & $0.911\,(0.047)$  \\
         & $1.5$ & \cmark & $0.924\,(0.042)$\\
         \midrule
         \multirow{7}{*}{{\large Barlow Twins}} & $0$ & \xmark & $0.905\,(0.061)$  \\
         & $0.5$ & \xmark & $0.912\,(0.048)$  \\
         & $0.5$ & \cmark & $0.904\,(0.052)$  \\
         & $1$ & \xmark & $0.914\,(0.057)$  \\
         & $1$ & \cmark & $0.911\,(0.052)$  \\
         & $1.5$ & \xmark & $0.921\,(0.054)$ \\
         & $1.5$ & \cmark & $0.910\,(0.041)$ \\
         \midrule
         \multirow{7}{*}{{\large VICReg}}  & $0$ & \xmark & $0.917 (0.046)$   \\
         & $0.5$ & \xmark & $0.924\, (0.044)$  \\
         & $0.5$ & \cmark & $0.912\, (0.050)$ \\
         & $1$ & \xmark & $0.925\, (0.052)$  \\
         & $1$ & \cmark & $0.930\, (0.046)$  \\
         & $1.5$ & \xmark & $0.913\, (0.045)$ \\
         & $1.5$ & \cmark & $0.918\, (0.035)$  \\
         \midrule
        None (ImageNet-pretrained) & & & $0.908\,(0.043)$ \\
        None (random initialization) & & & $0.881\,(0.050)$ \\
        \bottomrule
    \end{tabular}
    \caption{
    \normalsize Mean $5-$fold cross validation test accuracy for ResNet18 classifiers pretrained on the Butterfly dataset and fine-tuned on the POCUS dataset.
    }
    \label{tab:pocus-results}
\end{table}

\begin{table}[h!]
    \small
    \centering
    \begin{tabular}{c@{\hskip 0.6in}ccc@{\hskip 0.6in}ccc}
        \toprule
        & \multicolumn{3}{c}{{\large \tt AB}} & \multicolumn{3}{c}{{\large \tt LS}} \\[5pt]
         Pretraining Method & $\delta_t$ & SW & Test AUC & $\delta_x$ & SW & Test AUC \\
         \midrule
         \multirow{7}{*}{{\large SimCLR}} & $0$ & \xmark & $0.952$ & 0 & \xmark & $0.863$  \\
         & $0.5$ & \xmark & $0.962$ & $5$ & \xmark &  $0.867$ \\
         & $0.5$ & \cmark & $0.961$ & $5$ & \cmark & $0.874$ \\
         & $1$ & \xmark & $0.964$ & $10$ & \xmark & $0.865$ \\
         & $1$ & \cmark & $0.963$ & $10$ & \cmark & $0.872$ \\
         & $1.5$ & \xmark & $0.958$ & $15$ & \xmark & $0.864$ \\
         & $1.5$ & \cmark & $0.959$ & $15$ & \cmark & $0.860$ \\
         \midrule
         \multirow{7}{*}{{\large Barlow Twins}} & $0$ & \xmark & $0.903$ & $0$ & \xmark & $0.818$ \\
         & $0.5$ & \xmark & $0.922$ & $5$ & \xmark &  $0.832$ \\
         & $0.5$ & \cmark & $0.923$ & $5$ & \cmark & $0.853$ \\
         & $1$ & \xmark & $0.914$ & $10$ & \xmark & $0.834$ \\
         & $1$ & \cmark & $0.915$ & $10$ & \cmark & $0.849$ \\
         & $1.5$ & \xmark & $0.906$ & $15$ & \xmark & $0.838$ \\
         & $1.5$ & \cmark & $0.905$ & $15$ & \cmark & $0.856$ \\
         \midrule
         \multirow{7}{*}{{\large VICReg}}  & $0$ & \xmark & $0.938$ & 0 & \xmark & $0.830$  \\
         & $0.5$ & \xmark & $0.912$ & $5$ & \xmark &  $0.853$ \\
         & $0.5$ & \cmark & $0.921$ & $5$ & \cmark & $0.829$ \\
         & $1$ & \xmark & $0.907$ & $10$ & \xmark & $0.843$ \\
         & $1$ & \cmark & $0.904$ & $10$ & \cmark & $0.838$ \\
         & $1.5$ & \xmark & $0.916$ & $15$ & \xmark & $0.834$ \\
         & $1.5$ & \cmark & $0.9135$ & $15$ & \cmark & $0.837$ \\
         \midrule
        None (ImageNet-pretrained) & & & $0.906$ & & & $0.857$ \\
        None (random initialization) & & & $0.604$ & & & $0.516$ \\
        \bottomrule
    \end{tabular}
    \caption{
    \normalsize ParenchymalLUS test AUC for linear classifiers trained on the {\tt AB} and {\tt LS} tasks when trained on all examples in the training set.
    }
    \label{tab:linear-results}
\end{table}

\begin{table}[h!]
    \small
    \centering
    \begin{tabular}{c@{\hskip 0.6in}ccc@{\hskip 0.6in}ccc}
        \toprule
        & \multicolumn{3}{c}{{\large \tt AB}} & \multicolumn{3}{c}{{\large \tt LS}} \\[5pt]
         Pretraining Method & $\delta_t$ & SW & Test AUC & $\delta_x$ & SW & Test AUC \\
         \midrule
         \multirow{7}{*}{{\large SimCLR}} & $0$ & \xmark & $0.977$ & 0 & \xmark & $0.863$  \\
         & $0.5$ & \xmark & $0.977$ & $5$ & \xmark &  $0.893$ \\
         & $0.5$ & \cmark & $0.978$ & $5$ & \cmark & $0.893$ \\
         & $1$ & \xmark & $0.977$ & $10$ & \xmark & $0.897$ \\
         & $1$ & \cmark & $0.979$ & $10$ & \cmark & $0.898$ \\
         & $1.5$ & \xmark & $0.980$ & $15$ & \xmark & $0.903$ \\
         & $1.5$ & \cmark & $0.980$ & $15$ & \cmark & $0.898$ \\
         \midrule
         \multirow{7}{*}{{\large Barlow Twins}} & $0$ & \xmark & $0.969$ & $0$ & \xmark & $0.818$ \\
         & $0.5$ & \xmark & $0.968$ & $5$ & \xmark &  $0.870$ \\
         & $0.5$ & \cmark & $0.965$ & $5$ & \cmark & $0.882$ \\
         & $1$ & \xmark & $0.966$ & $10$ & \xmark & $0.863$ \\
         & $1$ & \cmark & $0.964$ & $10$ & \cmark & $0.879$ \\
         & $1.5$ & \xmark & $0.965$ & $15$ & \xmark & $0.887$ \\
         & $1.5$ & \cmark & $0.966$ & $15$ & \cmark & $0.886$ \\
         \midrule
         \multirow{7}{*}{{\large VICReg}}  & $0$ & \xmark & $0.971$ & 0 & \xmark & $0.830$  \\
         & $0.5$ & \xmark & $0.965$ & $5$ & \xmark &  $0.861$ \\
         & $0.5$ & \cmark & $0.963$ & $5$ & \cmark & $0.862$ \\
         & $1$ & \xmark & $0.962$ & $10$ & \xmark & $0.856$ \\
         & $1$ & \cmark & $0.966$ & $10$ & \cmark & $0.859$ \\
         & $1.5$ & \xmark & $0.969$ & $15$ & \xmark & $0.856$ \\
         & $1.5$ & \cmark & $0.967$ & $15$ & \cmark & $0.861$ \\
         \midrule
        None (ImageNet-pretrained) & & & $0.973$ & & & $0.898$ \\
        None (random initialization) & & & $0.954$ & & & $0.790$ \\
        \bottomrule
    \end{tabular}
    \caption{
    \normalsize ParenchymalLUS test AUC for classifiers fine-tuned on the {\tt AB} and {\tt LS} tasks when trained on all examples in the training set.
    }
    \label{tab:fine-tune-results}
\end{table}

\end{appendix}

\bibliographystyle{unsrt}  
\bibliography{references}

\end{document}